\documentclass[aps,prd,twocolumn,superscriptaddress]{revtex4-2}

\usepackage{amsmath,mathtools,amsthm,amssymb,pifont}
\usepackage[utf8]{inputenc}
\usepackage{graphicx}
\usepackage{caption}
\usepackage{ragged2e}
\usepackage{subcaption} 
\usepackage[american]{babel}
\usepackage{graphicx,xcolor,bbold,titlesec}
\usepackage{makecell}
\usepackage{braket}
\usepackage{MnSymbol}
\usepackage{tabularx}

\usepackage{xcolor}

\usepackage[colorlinks,
bookmarksopen,
bookmarksnumbered,
citecolor=red,
linkcolor=red,
pdfstartview=false,
urlcolor=red]{hyperref}
\usepackage{tikz,ifthen}
\usepackage{bbold}
\usepackage{tikz-network}
\usetikzlibrary{patterns,decorations.pathreplacing,calligraphy}
\usepackage{orcidlink}
\usepackage{color}
\usepackage{MnSymbol}

\begin{document}

\title{Exact stabilizer scars in two-dimensional $U(1)$ lattice gauge theory}
\author{Sabhyata Gupta~\orcidlink{0009-0003-3551-8651}}
\email{sabhyata.gupta@itp.uni-hannover.de}
\affiliation{Institut f\"ur Theoretische Physik, Leibniz Universit\"at Hannover, Appelstrasse 2, 30167, Germany }

\author{Piotr Sierant~\orcidlink{}}
\email{piotr.sierant@bsc.es}
\affiliation{Barcelona Supercomputing Center Plaça Eusebi G\"uell, 1-3 08034, Barcelona, Spain}

\author{Luis Santos~\orcidlink{}}
\email{santos@itp.uni-hannover.de}
\affiliation{Institut f\"ur Theoretische Physik, Leibniz Universit\"at Hannover, Appelstrasse 2, 30167, Germany }

\author{Paolo Stornati~\orcidlink{0000-0003-4708-9340}}
\email{paolo.stornati@bsc.es}
\affiliation{Barcelona Supercomputing Center Plaça Eusebi G\"uell, 1-3 08034, Barcelona, Spain}

\begin{abstract}
The complexity of highly excited eigenstates is a central theme in nonequilibrium many-body physics, underpining questions of thermalization, classical simulability, and quantum information structure. In this work, considering the paradigmatic Rokhsar–Kivelson model, we connect quantum many-body scarring in Abelian lattice gauge theories to an emergent stabilizer structure. We identify a distinct class of scarred eigenstates, termed sublattice scars, originating from gauge-invariant zero modes that form exact stabilizer states. Remarkably, although the underlying Hamiltonian is not a stabilizer Hamiltonian, its eigenspectrum intrinsically hosts exact stabilizer eigenstates. These sublattice scars exhibit vanishing stabilizer Rényi entropy together with finite, highly structured entanglement, enabling efficient classical simulation. Exploiting their stabilizer structure, we construct explicit Clifford circuits that prepare these states in a two-dimensional lattice gauge model. Our results demonstrate that the scarred subspace of the Rokhsar–Kivelson spectrum forms an intrinsic stabilizer manifold, revealing a direct connection between stabilizer quantum information, lattice gauge constraints, and quantum many-body scarring.
\end{abstract}

\maketitle

\section{Introduction}

Isolated quantum many-body systems are expected to self-thermalize under unitary dynamics, a behavior encapsulated by the eigenstate thermalization hypothesis (ETH), which asserts that individual high-energy eigenstates reproduce thermal expectation values of local observables~\cite{Deutsch1991,Srednicki1994,DAlessio2016}. An important exception comes from quantum many-body scars~\cite{Turner2018,Serbyn2021}, a class of highly excited eigenstates that weakly violate ETH and give rise to nonthermal dynamics despite being embedded in an otherwise thermal spectrum. Scarred eigenstates exhibit anomalously low entanglement and lead to long-lived coherent revivals when the system is initialized in a special class of product states~\cite{Bernien2017,Ho2019}.

Constrained quantum systems provide a natural setting for this persistent coherent dynamics~\cite{Chandran2023}. In particular, models governed by local conservation rules such as Gauss's law in lattice gauge theories (LGTs) restrict the Hilbert space to a physical subspace, which can harbor nonergodic dynamics and long-lived states~\cite{Surace2020,ScarsPXPPhysRevLett.134.160401,Halimeh2023robustquantummany,banerjee2021scars,QMBS1dPhysRevLett.132.230403,biswas2022scars,ExactQMBSPhysRevB.108.195133,lan2017eth,PhysRevD.110.094506,ScarsSchwingerPhysRevB.107.205112,sublatticescars2024, Brenes18, Szoldra22}. Recent studies have shown that constrained Hilbert spaces can protect special nonthermal eigenstates through symmetry-based mechanisms, Hilbert-space fragmentation~\cite{Sala2020,Khemani2020}, or entanglement bottlenecks~\cite{Moudgalya2022}.

\begin{figure}[t!]
    \centering
    \includegraphics[width=1\linewidth]{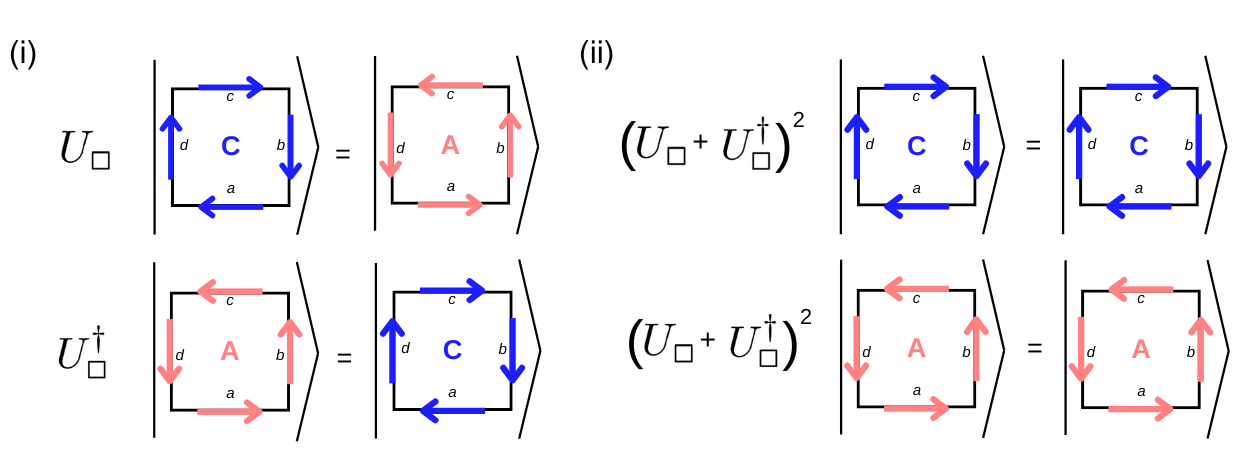}
    \caption{\justifying Schematic representation of the RK Hamiltonian: action of $(i)$ $O_{\mathrm{kin}}$ $(ii)$ $O_{\mathrm{pot}}$ operators on active/flippable plaquettes.}
    \label{fig:plaquette}
\end{figure}

The Rokhsar-Kivelson (RK) model~\cite{rokhsar1988superconductivity} provides a paradigmatic example of a constrained, gauge-invariant quantum system. Originally introduced in the context of quantum dimer models, the RK Hamiltonian features plaquette flip terms (see Fig.~\ref{fig:plaquette}) acting on a background of hardcore dimers obeying local dimer constraints, a variant of Gauss's law. The model supports a solvable spin-liquid ground state and has been studied in connection to resonating valence bond phases, topological order, and lattice gauge theories~\cite{Moessner2001,Banerjee2014}.

Recent works have shown classes of stabilizer quantum many body scarred states in $\mathbb{Z}2$ LGT, and other b-local Hamiltonians~\cite{Stabilizerscars,dooley2026parenthamiltoniansstabilizerquantum}.  Stabilizer Hamiltonians, e.g.~\cite{KITAEV20032}, by definition, possess stabilizer eigenstates that are fixed by a commuting set of Pauli operators. 
However, such Hamiltonians are typically artificial constructions composed of mutually commuting projector terms and do not represent the type of local, physical models. Beyond exact stabilizer Hamiltonians, recent theoretical developments have established that “doped” stabilizer states, generalizations of exact stabilizer states with limited non-Clifford content, appear naturally as eigenstates of perturbed many-body Hamiltonians, enabling stabilizer techniques to be applied in highly entangled regimes~\cite{PhysRevA.110.062427}.
In contrast, the RK model is not a stabilizer Hamiltonian, as its kinetic and potential terms do not commute. Yet, as shown below, its spectrum nonetheless contains exact stabilizer eigenstates.



In this work, we uncover and characterize a class of eigenstates in the two-dimensional RK model that are both scarred and exact stabilizer states. These stabilizer scars possess zero stabilizer Rényi entropy (SRE)~\cite{leone2022stabilizer, Leone24} and limited bipartite entanglement. We analytically verify their canonical stabilizer form and show that they are invariant under a commuting set of physical Pauli operators. These eigenstates violate ETH, are exact stabilizer states, and hence lie within the class of classically simulable quantum many-body states~\cite{StabilizerCircuitsPhysRevA.70.052328}. We further provide explicit Clifford circuits that prepare these stabilizer scar states efficiently, showing that they are accessible on near-term quantum devices. 

Our results demonstrate that the scarred subspace formed by the sublattice singlet states of the RK model 
constitutes an exact stabilizer manifold within the gauge-invariant Hilbert space. 
This establishes a direct correspondence between sublattice scars and stabilizer-protected subspaces, bridging quantum many-body scarring, lattice-gauge constraints, and the theory of stabilizer states.

\section{Model}
\label{sec:model}
The RK model arises in the theory of quantum dimer~\cite{rokhsar1988superconductivity} and spin-ice~\cite{Hermele04, Shannon12} systems. It captures the essential ingredients of lattice gauge theories, namely local ring-exchange terms and Gauss-law constraints, giving rise to non trivial phenomenon, like cofinement~\cite{Banerjee_2013} and non trivial phases of matter~\cite{Stornati_2023}. We focus here on the spin-$\frac{1}{2}$ formulation of the RK model, in which each link of the lattice hosts a spin degree of freedom. The Hamiltonian is expressed in terms of plaquette operators:
\begin{equation}
\begin{aligned}
H_{\mathrm{RK}} &= \mathcal{O}_{\mathrm{kin}} + \lambda \mathcal{O}_{\mathrm{pot}} \\
&= -\sum_{\square} \mathcal{O}_{\mathrm{kin},\square} + \lambda \sum_{\square} \mathcal{O}_{\mathrm{pot},\square} \\
&= -\sum_{\square} \left(U_{\square} + U_{\square}^\dagger \right) + \lambda \sum_{\square} \left(U_{\square} + U_{\square}^\dagger \right)^2,
\end{aligned}
\end{equation}
where the sums run over all plaquettes. 
The plaquette operator $U_{\square} = S_a^{+} S_b^{+} S_c^{-} S_d^{-}$ 
acts on the four spins $(a,b,c,d)$ around the plaquette $\square$, flipping a 
clockwise configuration $\ket{C}$ into the corresponding 
anticlockwise configuration $\ket{A}$, i.e.\ 
$U_{\square}\ket{A} = \ket{C}$ and 
$U_{\square}^\dagger\ket{C} = \ket{A}$, see Fig.~\ref{fig:plaquette}. 
Any other spin configuration is inactive, i.e. non-flippable, and is annihilated by $U_{\square}$ and $U_{\square}^\dagger$. 
Hence, the kinetic term $\mathcal{O}_{\mathrm{kin}}$ acts only on the 
active, i.e. flippable, plaquettes, while the potential term 
$\mathcal{O}_{\mathrm{pot}} = \sum_{\square} (U_{\square} + U_{\square}^\dagger)^2$ 
counts the number of active plaquettes in a given configuration.

The local Hilbert space is constrained by a Gauss law. The  local Gauss law operator at each vertex $r$ is
\begin{equation}
G_r = \sum_{\substack{\mu \in \{ \hat{x}, \hat{y} \}}} \left( S^z_{r,\hat{\mu}} - S^z_{r - \hat{\mu}, \hat{\mu}} \right),
\end{equation}
where $S^z_{r,\hat{\mu}}$ denotes the $z$-component of the spin on the link emanating from site $r$ in the $\hat{\mu}$ direction. 
The RK Hamiltonian satisfies 
$[H_{\mathrm{RK}}, G_r] = 0$, 
ensuring that the dynamics remains confined to the gauge-invariant subspace. In this work, we restrict to the charge neutral sector, in which the physical Hilbert space is defined by the set of gauge-invariant states that satisfy the local constraint 
$G_r \ket{\psi} = 0,   \forall r$.

\section{Sublattice scars}

In this model, a class of anomalous eigenstates called sublattice scars, has been identified in Ref.~\cite{PhysRevLett.126.220601,sublatticescars2024}. 
These states are characterized by integer-valued eigenvalues of both the kinetic and potential operators, 
\(\mathcal{O}_{\mathrm{kin}}\) and \(\mathcal{O}_{\mathrm{pot}}\), within the constrained Hilbert space.
Specifically, dividing the two-dimensional lattice into two sublattices in a checker-board configuration, a sublattice scar \(|\psi_s\rangle\) satisfies \(\mathcal{O}_{\mathrm{pot},\square}|\psi_s\rangle = |\psi_s\rangle\) for all plaquettes on one sublattice and
\(\mathcal{O}_{\mathrm{pot},\square}|\psi_s\rangle = 0\) on the complementary sublattice, 
while being an eigenstate of \(\mathcal{O}_{\mathrm{kin}}\) with integer eigenvalue 
\(\mathcal{O}_{\mathrm{kin}} \lvert \psi_s \rangle = n \lvert \psi_s \rangle,\; n \in \{0, \pm 2\}\). By definition, such states remain exact eigenstates of the full Hamiltonian $H = \mathcal{O}_{\mathrm{kin}} + \lambda \mathcal{O}_{\mathrm{pot}}$ for any value of the coupling $\lambda$, implying that they are isolated from the ergodic continuum and do not hybridize with nearby eigenstates as $\lambda$ is varied.  Physically, the integer spectrum of \(\mathcal{O}_{\mathrm{kin}}\) signals a coherent plaquette-flip pattern localized on one sublattice, resulting in a robust athermal subspace embedded within the otherwise ergodic spectrum. While there are several other classes of scars identified within the RK model, in this work we focus on the sublattice scars \cite{Banerjee2014,banerjee2021scars}.


\section{Complexity markers}

\subsection{Stabilizer structure}
\label{Stabilizer diagnostics}
To quantify the magic resources, i.e. non-stabilizerness, of a given state $|\psi \rangle$,  we compute  the stabilizer Rényi entropy (SRE)~\cite{leone2022stabilizer}. The SRE of order $n$ is defined as
\begin{equation}
    M_n(|\psi\rangle) = \frac{1}{1 - n} \log \left( \sum_{P \in \mathcal{P}_N} \frac{|\langle \psi | P | \psi \rangle|^{2n}}{2^N} \right),
\end{equation}
where the sum runs over the full \(N\)-qubit Pauli group \( \mathcal{P}_N \), and the logarithm is taken in the natural base. For \(n = 2\)~\footnote{Fully analogous properties hold for any integer $n>2$.}.
The SRE is a monotone in magic resource theory of pure states~\cite{Leone24}, and quantifies deviation of $\ket{\psi}$ from being a stabilizer state. 
Stabilizer states yield \(M_2 = 0\), while any state requiring non-Clifford resources for preparation yields a strictly positive value of $M_2$. 

Although $M_2$ may be computed exactly for small system sizes, 
its evaluation becomes computationally challenging for larger systems, as it requires calculating the expectation values of all $4^N$ Pauli operators~\cite{Sierant26computingSRE}.
To avoid this difficulty, we employ multifractal flatness~$\tilde{\mathcal{F}}(\ket{\psi})$ \cite{MultifractalFlatness2023}, whose evaluation is significantly less demanding for states written in the computational basis, which in our case is the Fock-like basis of gauge-invariant spin configurations $\{|{\sigma \rangle}\}$. Multifractal flatness is defined via the participation probabilities $p_\sigma = |\langle \sigma | \psi \rangle|^2$:
\begin{equation}
\tilde{\mathcal{F}}(\ket{\psi}) =
\sum_{\sigma} p_\sigma^{3}
 - \left( \sum_{\sigma} p_\sigma^{2} \right)^{2}.
\end{equation}

The multifractal flatness measures the deviation of the participation distribution $\{p_\sigma\}$ from a completely uniform spread, vanishing in the limit where all $p_\sigma$ are equal. If $\ket{\psi}$ is a stabilizer state, the multifractal flatness is vanishing~\cite{Sierant22multi}, $\tilde{\mathcal{F}}(\ket{\psi}) =0$.

The converse, however, is not generally true; a flat eigenstate corresponds to a stabilizer state only if its wavefunction takes the canonical form~\cite{Dehaene03, montanaro2017learningstabilizerstatesbell}
\begin{equation}
|\psi\rangle = \frac{1}{\sqrt{|A|}} 
\sum_{x \in A} i^{\ell(x)} (-1)^{q(x)} |x\rangle,
\label{eq:canonical_form}
\end{equation}
where $A$ is an affine subspace of $\mathbb{F}_2^n$, and 
$\ell, q : \{0,1\}^n \rightarrow \{0,1\}$ are linear and quadratic polynomials, respectively, over the finite field $\mathbb{F}_2$. 
This implies that a flat state is uniformly supported on an affine subspace $A$ of the computational basis, with relative phases determined by linear and quadratic functions. Although the affine-phase structure is strictly required for a state to be a stabilizer state, the condition $\tilde{\mathcal{F}}(|\psi\rangle) = 0$ serves as a practical filter to identify candidate stabilizer states without the need to compute the SRE.

In the following, we employ these criteria, either $M_2 = 0$ or the verification of the canonical stabilizer representation when monitoring the existence of stabilizer states. 


\subsection{Signatures of scar-behavior}

The entanglement entropy~(EE)~\cite{Horodecki09quantum, Amico08entanglement} is a key diagnostic to distinguish thermal and scarred eigenstates: 
thermal states obey the volume-law scaling predicted by ETH, while scarred or stabilizer states exhibit lower entanglement. Splitting the system at half the system length, and obtaining the reduced density matrix $\rho_A$ in one of the halves, we evaluate the EE, $S_{\mathrm{vN}} = -\mathrm{Tr}(\rho_A \log \rho_A)$, where the logarithm is taken in natural base. For stabilizer states, EE is strictly quantized to integer multiples of $\log(2)$~\cite{Hamma05}, corresponding to a perfectly flat entanglement spectrum~\cite{Tirrito24}.

Furthermore, as discussed above, the sublattice scarred eigenstates satisfy 
 $\mathcal{O}_{\mathrm{kin}}|\psi_s\rangle = n|\psi_s\rangle$ with $n = 0, \pm 2$, and $\mathcal{O}_{\mathrm{pot},\square}|\psi_s\rangle = |\psi_s\rangle$ on all plaquettes of one sublattice and $0$ on the complementary one. 
These extremal eigenvalues correspond to maximally ordered plaquette configurations that contrast sharply with homogeneous thermal averages ($\langle \mathcal{O}_{\mathrm{pot}}\rangle_{\mathrm{th}}\!\approx\!1/2$) predicted by ETH. We thus compute the expectation values of the kinetic and potential operators, $\mathcal{O}_{\mathrm{kin}}$ and $\mathcal{O}_{\mathrm{pot}}$, respectively, to characterize scarring behaviour.


\subsection{Degeneracy and basis dependence}

The RK Hamiltonian $H_{\mathrm{RK}} = \mathcal{O}_{\mathrm{kin}} + \lambda\,\mathcal{O}_{\mathrm{pot}}$ possesses an extensively degenerate eigenspectrum due to the local gauge constraints and the fact that 
$\mathcal{O}_{\mathrm{kin}}$ and $\mathcal{O}_{\mathrm{pot}}$ do not commute with each other, and therefore also not with $H_{\mathrm{RK}}$ for any $\lambda$. 
As a result, exact numerical diagonalization yields a specific orthonormal eigenbasis within each degenerate energy manifold, but the individual eigenvectors in that manifold are not uniquely defined. 
Because $M_2$ is explicitly basis-dependent, performing orthogonal rotations within these degenerate subspaces can reveal different stabilizer structures. 
We exploit this freedom to systematically engineer an orthonormal basis in which certain linear combinations of degenerate eigenstates achieve $M_2 = 0$.
Such basis engineering enables the identification of hidden stabilizer submanifolds embedded in the 
highly degenerate spectrum of the RK model.
To systematically uncover such states, we perform a basis rotation within each degenerate energy sector guided by the canonical form of stabilizer states. For a fixed eigenvalue $E$, we first identify the degenerate subspace $\mathcal{H}_E = \mathrm{span}\{|\psi_\alpha\rangle\}$. Within this subspace, we search for linear combinations $|\phi\rangle = \sum_\alpha c_\alpha |\psi_\alpha\rangle$, 
that admit a canonical stabilizer representation~\eqref{eq:canonical_form}. Such canonical patterns characterize stabilizer states up to Clifford transformations, and hence provide a natural variational manifold for identifying states with $M_2 = 0$.

The operators $\mathcal{O}_{\mathrm{kin}}$ and $\mathcal{O}_{\mathrm{pot}}$ play a crucial role in this construction. Although they do not commute, their expectation values on classical plaquette configurations are integer valued revealing scar nature. We therefore evaluate $\langle \mathcal{O}_{\mathrm{kin}} \rangle$ and $\langle \mathcal{O}_{\mathrm{pot}} \rangle$ on candidate superpositions and retain only those states for which both quantities take sharply defined integer values. Empirically, this criterion strongly correlates with the stabilizer structure and selects states that are simultaneously structured with respect to plaquette flips and potential energy contributions.


Having identified in that way a set of candidate states within $\mathcal{H}_E$, we orthonormalize them using a Gram-Schmidt procedure restricted to the degenerate subspace. The remaining orthogonal complement is arbitrarily completed to produce a complete orthonormal basis of $\mathcal{H}_E$, with stabilizer scar states appearing explicitly as basis vectors.


\section{Stabilizer sublattice scars}
\label{sec:results}

We have evaluated the existence of stabilizer sublattice scars in RK models of different system sizes, considering periodic boundary conditions~(PBC).  Stabilizer sublattice scars appear systematically across all system sizes considered, as summarized in Table~\ref{tab:system_summary}.

\begin{figure}[b!]
    \centering
    \includegraphics[width=0.85\linewidth]{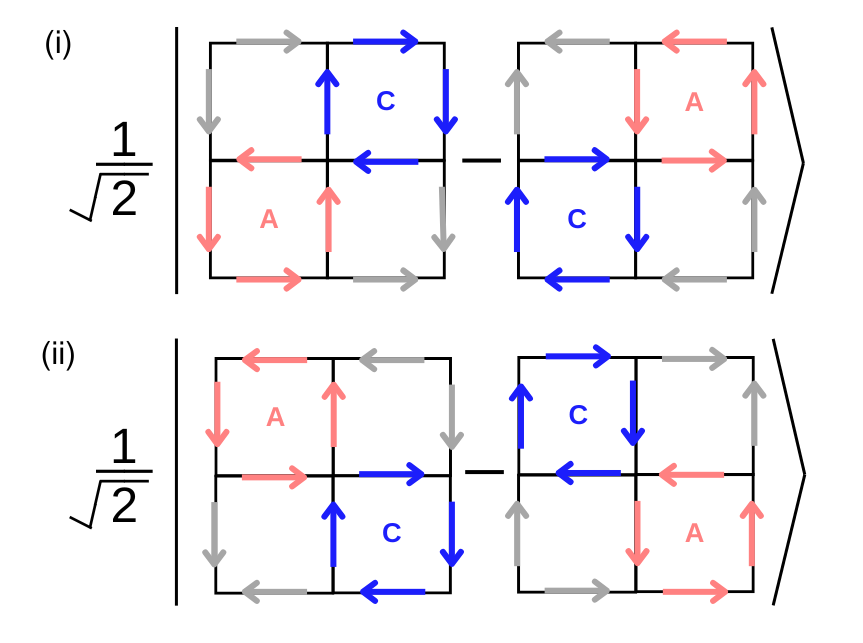}
    \caption{\justifying $(i) -(ii)$ Stabilizer sublattice scars in a $2\times2$ RK model with PBC in the energy sector $E=2$. These states possess $M_2 = 0$, $S_{\mathrm{vN}} = \ln 2$ and $\mathcal{O}_{\mathrm{kin}} = 0$, and exhibit $\mathcal{O}_{\mathrm{pot},\square} = 1$ on plaquettes belonging to one sublattice and $0$ on the complementary one.}
    \label{fig:2x2pbc}
\end{figure}

We compute the SRE exactly for systems with $2 \times 2$ and $4 \times 2$ plaquettes. For larger systems we employ the alternative diagnostics as described in Sec.~\ref{Stabilizer diagnostics}. For the $2 \times 2$, we identify $6$ stabilizer states with $M_2 = 0$. Out of these, 4  correspond to gauge-invariant product configurations, or Fock-like states, in energy sector $E=0$ and exhibit zero entanglement entropy. They are therefore identified as trivial stabilizer states. 
We find 2 eigenstates in the energy sector $E=2$, with $M_2 = 0$ and bipartite entanglement entropy $S_{\mathrm{vN}} = \ln 2$. These states possess $\mathcal{O}_{\mathrm{kin}} = 0$ and exhibit $\mathcal{O}_{\mathrm{pot},\square} = 1$ on all plaquettes belonging to one sublattice and $0$ on the complementary one, corresponding to the characteristic sublattice-scar configuration, see Fig.~\ref{fig:2x2pbc}. These states are hence examples of 
stabilizer sublattice scars.



\begin{table}[t!]
\centering
\small
\captionsetup{justification=justified}
\setlength{\tabcolsep}{3pt}
\renewcommand{\arraystretch}{1.1}
\begin{tabular}{@{}lccc@{}}
\hline
\textbf{System} & 
\makecell[c]{\textbf{No. of}\\\textbf{spins}} & 
\makecell[c]{\textbf{No. of}\\\textbf{physical states}} & 
\makecell[c]{\textbf{No. of}\\\textbf{stabilizer scars}} \\
\hline
$2\times2$ & 8  & 18      & 2 \\
$4\times2$ & 16 & 114     & 2 \\
$6\times2$ & 24 & 858     & 2 \\
$4\times4$ & 32 & 2{,}970  & 8\\
\hline
\end{tabular}
\caption{\justifying Summary of system sizes considered $L_x\times L_y$, with $L_{x,y}$ the number of plaquettes along $x$ and $y$. The table 
details the number of spins involved, the number physical states with zero charge, and the number of identified stabilizer sublattice scars.}
\label{tab:system_summary}
\end{table}




\begin{figure}[t!]
    \centering
    \includegraphics[width=1\linewidth]{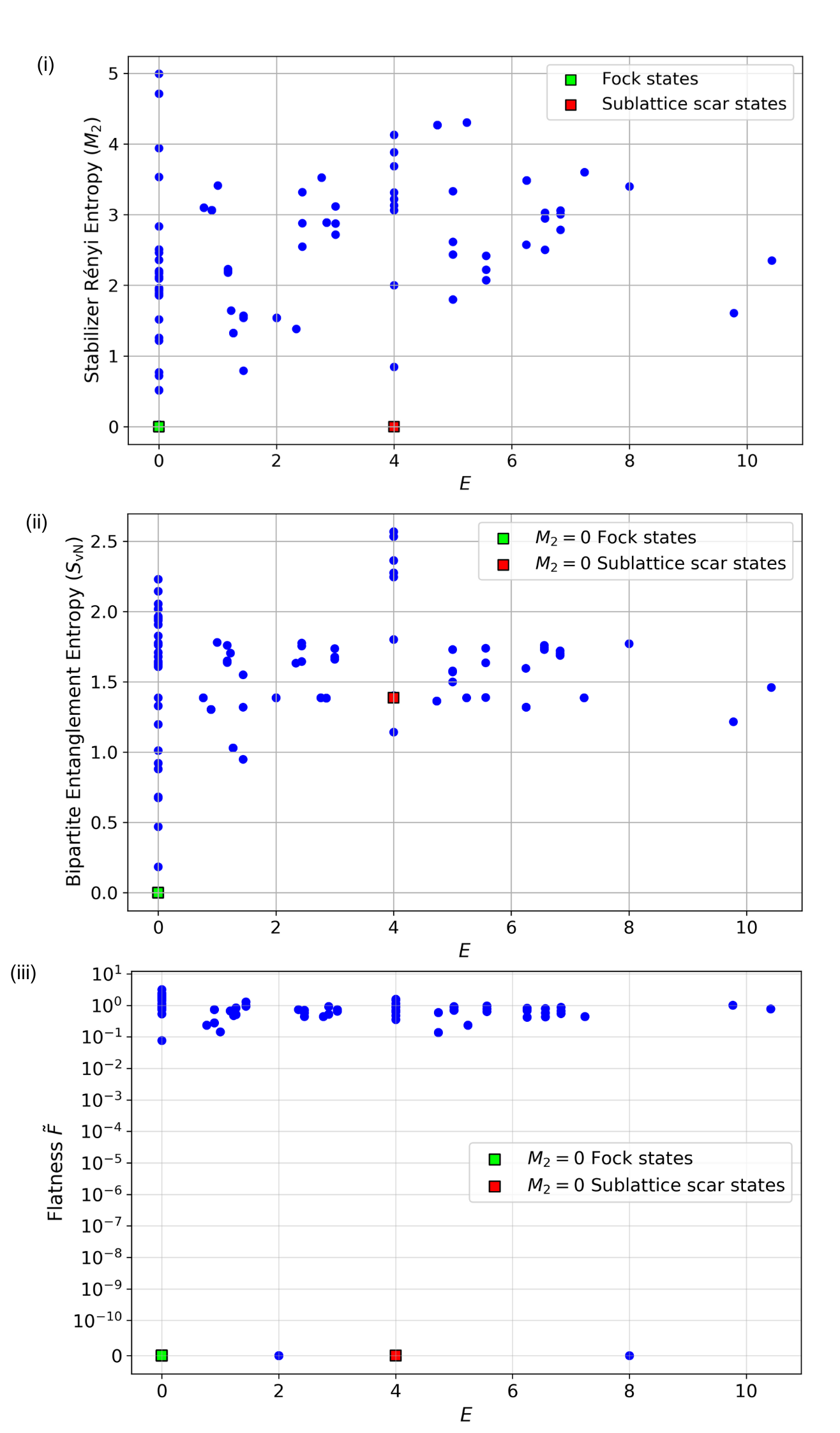}
    \caption{\justifying $(i)$ Stabilizer Rényi entropy $M_2$, $(ii)$ bipartite entanglement entropy $S_{vN}$, and $(iii)$ multifractal flatness~$\tilde F$ across the eigenspectrum for the $4\times2$ plaquette system with PBC. 
In $(i)$, two sublattice stabilizer scars at $E=4$ exhibit vanishing $M_2$, with trivial Fock stabilizer states marked in green and nontrivial sublattice stabilizer scars marked in red. 
In $(ii)$, the Fock states show $S_{\mathrm{vN}}=0$ owing to their product-state nature, while the sublattice stabilizer scars display finite entanglement $S_{\mathrm{vN}}=2\times \ln 2$. 
Panel $(iii)$ shows the multifractal flatness, where states at $E=2$ and $E=8$ appear flat but do not correspond to valid stabilizer states, consistent with the absence in $(i)$ of $M_2=0$ states at $E=2$ and $8$.
}
    \label{fig:4x2pbc}
\end{figure}


As shown in Figs.~\ref{fig:4x2pbc}, for the $4 \times 2$ system, we find 2 sublattice stabilizer scars in the energy sector $E = 4$, characterized by vanishing $M_2$ and bipartite entanglement entropy $S_{\mathrm{vN}} =2\times \ln 2$. While the stabilizer Rényi entropy is computed exactly, see Fig.~\ref{fig:4x2pbc} $(i)$, we additionally evaluate the multifractal flatness to benchmark the canonical-form analysis. A state may exhibit uniform amplitude magnitudes in the computational basis yet fail to satisfy the affine-support and quadratic-phase conditions required for the canonical form in Eq.~(\ref{eq:canonical_form}). Consequently, flat states observed at $E=2$ and $E=8$ in Fig.~\ref{fig:4x2pbc} $(iii)$ do not correspond to valid stabilizer manifolds, consistent with the absence of $M_2=0$ signatures in Fig.~\ref{fig:4x2pbc} $(i)$.

A similar analysis was carried out for larger system sizes. In lattice systems $L_x \times 2$, sublattice scars occur only as zero modes of ${\mathcal O}_{\mathrm{kin}}$.

\begin{figure}[t!]
    \centering
    \includegraphics[width=1\linewidth]{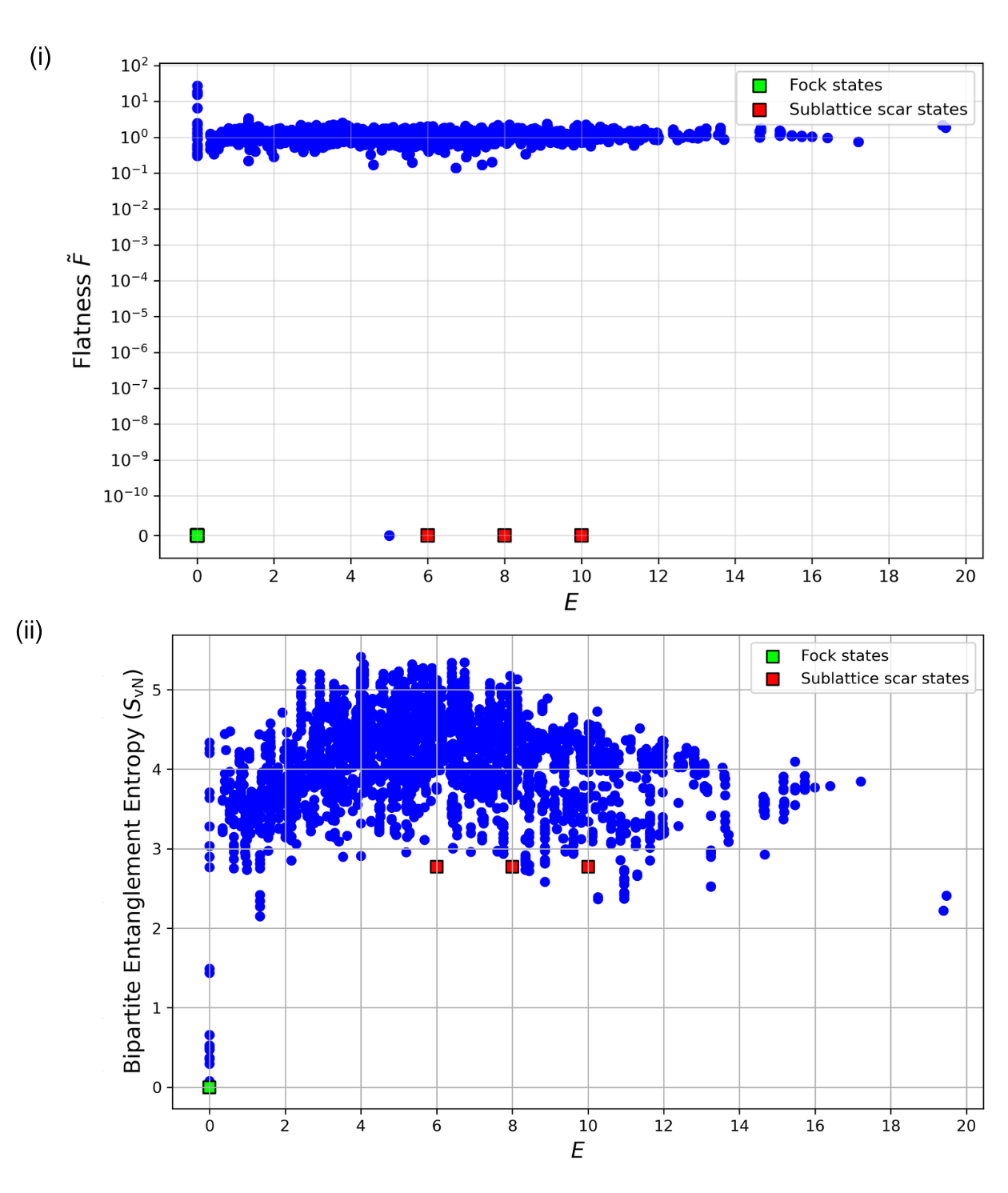}
    \caption{\justifying \justifying $(i)$ Multifractal flatness, $(ii)$ bipartite entanglement entropy, across the eigenspectrum for the $4\times4$ plaquette system with periodic boundary conditions.  }
    \label{fig:4x4observables}
\end{figure}

The situation is different for larger system sizes. For the largest system we have considered, $4\times4$, we observe $4$ sublattice stabilizer scars at energy $E=8$ which are zero modes of $\mathcal{O}_{\mathrm{kin}}$, but we also see 4 stabilizer sublattice scars with $\mathcal{O}_{\mathrm{kin}}|\psi_s\rangle = \pm 2|\psi_s\rangle$, $2$ at energies $E=6$ and other $2$ at $E=10$ as shown in Fig.~\ref{fig:4x4observables}. Although the number of sublattice scars has been reported to be higher in \cite{sublatticescars2024}, the canonicalization within the degenerate sector is not unique and yields a stabilizer structure under the given choice of canonicalization.  



\begin{figure}[t!]
    \centering
    \includegraphics[width=0.5\linewidth]{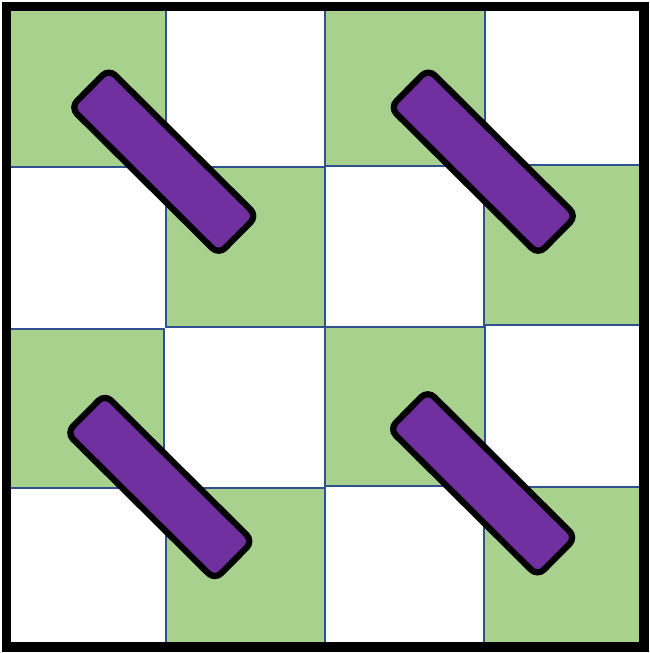}
    \caption{\justifying Sketch of the general structure of stabilizer sublattice scars. Green~(white) plaquettes are active~(inactive). The plaquettes linked by a purple rod formed a dimer in a singlet state.}
    \label{fig:dimers}
\end{figure}

The nature of stabilizer scar states with $\mathcal{O}_{\mathrm{kin}} = 0$ 
for any system size may be well understood by realizing that the flippable subspace $\mathcal{H}_{L,\square} 
=\mathrm{span}\{\ket{C},\ket{A}\}$ act as an effective two-dimensional
logical subspace on plaquette~$\square$. 
We may then introduce Pauli operators in that subspace: 
$X_{L,\square} \equiv \ket{C}\!\bra{A} + \ket{A}\!\bra{C}$ and 
$Z_{L,\square}\equiv \ket{C}\!\bra{C} - \ket{A}\!\bra{A} \equiv P_C - P_A$.
Plaquettes on the active sublattice are paired into dimers $\mathcal{D}=\{(p,q)\}$, see Fig.~\ref{fig:dimers},
while the complementary sublattice remains inactive. The local dimer state is the logical Bell singlet
\begin{equation}
\ket{\Psi^-}_{pq} = \frac{\ket{C}_p\ket{A}_q-\ket{A}_p\ket{C}_q}{\sqrt2}
\end{equation}
which satisfies the logical stabilizers
\begin{equation}
\begin{aligned}
(Z_{L,p}Z_{L,q})\ket{\Psi^-}_{pq} &= -\ket{\Psi^-}_{pq},\\
(X_{L,p}X_{L,q})\ket{\Psi^-}_{pq} &= -\ket{\Psi^-}_{pq}.
\end{aligned}
\label{eq:bell-stabs}
\end{equation}
The global sublattice singlet state is then:
\begin{equation}
\ket{\psi_{\mathrm{SS}}}=\bigotimes_{(p,q)\in\mathcal{D}}\ket{\Psi^-}_{pq}\ \otimes\ \ket{\mathrm{inactive}}.
\end{equation}
These states fulfill that $\mathcal{O}_{\mathrm{kin}}\ket{\psi_{\mathrm{SS}}}=0$, and 
$\mathcal{O}_{\mathrm{pot}}\ket{\psi_{\mathrm{SS}}}=M\,\ket{\psi_{\mathrm{SS}}}$, and hence 
$[\mathcal{O}_{\mathrm{kin}},\mathcal{O}_{\mathrm{pot}}]\ket{\psi_{\mathrm{SS}}}=0$, 
i.e. the scar lies in a common invariant subspace even though
$[\mathcal{O}_{\mathrm{kin}},\mathcal{O}_{\mathrm{pot}}]\neq 0$ on the full Hilbert space.
Thus, the sublattice short-singlet codespace is fixed by a stabilizer algebra $\mathcal{S}$,
while the logical operators generated by $X_{L,\square},Z_{L,\square}$ act within it.

Noting that ${\mathcal O}_{\mathrm{kin}}=\sum_\square X_{L,\square}$, 
it is straightforward to see that ${\mathcal O}_{\mathrm{kin}}|\psi_{SS}\rangle = 0$. Moreover, 
${\mathcal O}_{\mathrm{pot}}|\psi_{SS}\rangle = M|\psi_{SS}\rangle$, with $M$ equal to one half of the number of plaquettes. Furthermore, the states present Bell-pair entanglement ($S_{\mathrm{vN}}=\ln2$ per crossed dimer) with vanishing stabilizer Rényi entropy ($M_2=0$), and hence the states $\vert\psi_{SS}\rangle$ constitute a stabilizer sublattice scar embedded in the RK spectrum.



\section{Quantum Circuits to prepare stabilizer scars}
\label{sec:circuits}

\begin{figure}[t!]
    \centering
    \includegraphics[width=0.5\linewidth]{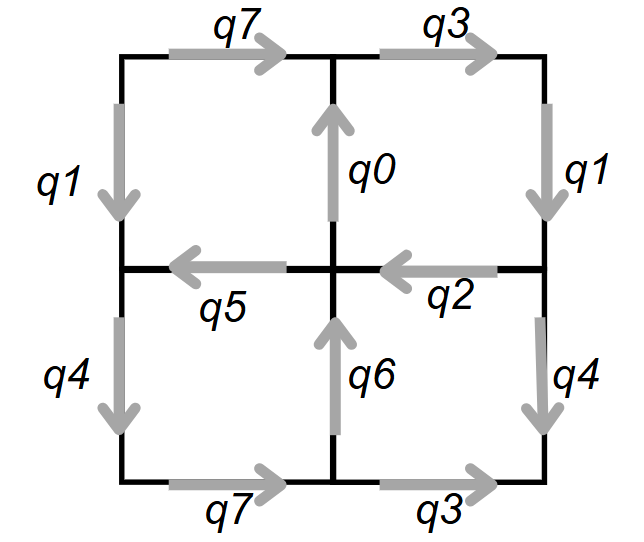}
    \caption{\justifying Qubit layout for $2\times2$ plaquettes with PBC.}
    \label{fig:qubit_layout}
\end{figure}

\begin{figure}[htpb]
    \centering
    \includegraphics[width=0.8\linewidth]{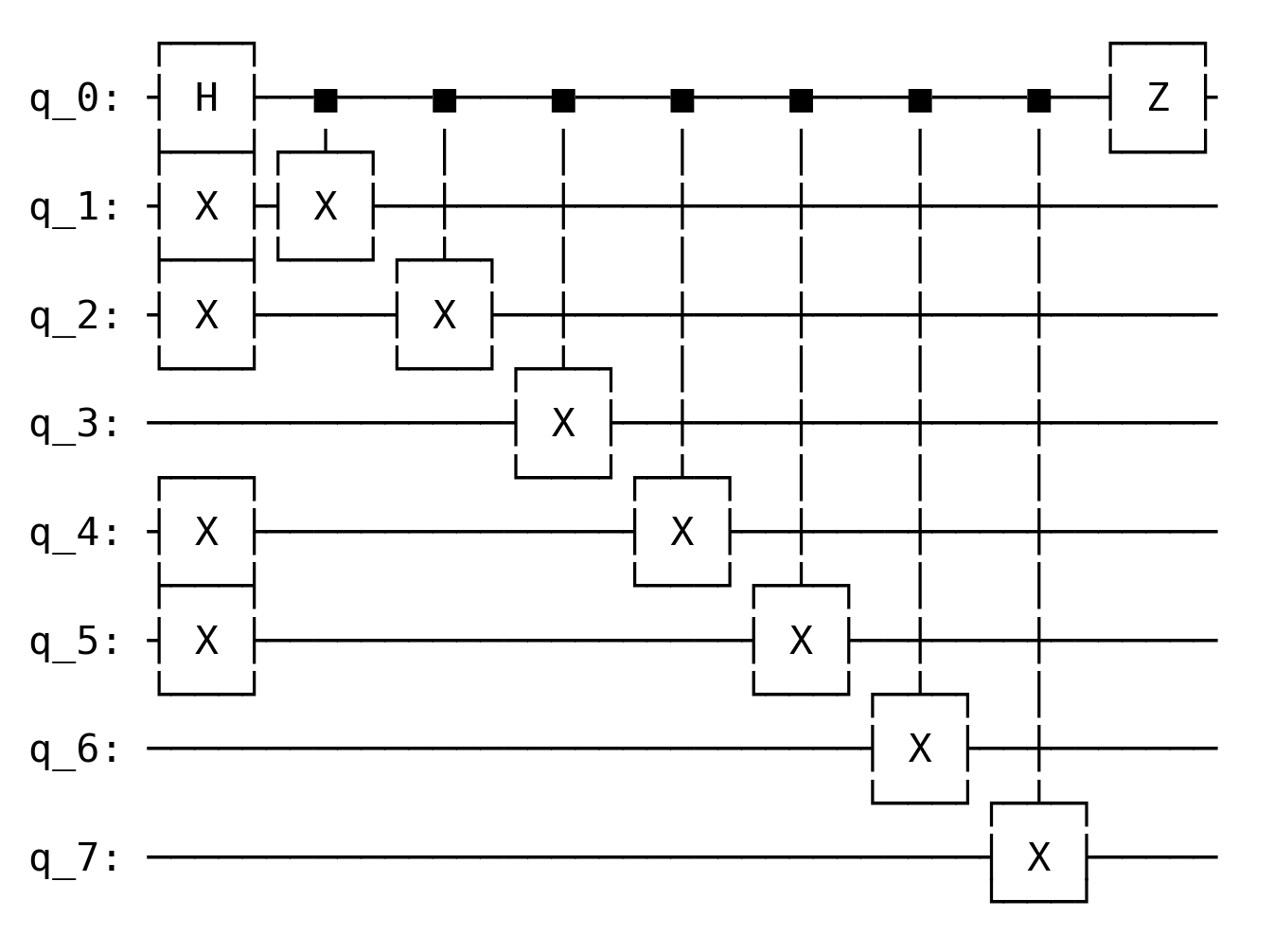}
    \caption{\justifying Quantum circuit for preparing a stabilizer scar for a $2\times2$ plaquette system with PBC.}
    \label{fig:circuit}
\end{figure}

We next outline the explicit Clifford circuit preparation for the minimal sublattice stabilizer scar configuration of Fig.~\ref{fig:2x2pbc}. This construction provides an experimentally accessible routine to initialize the non-thermal stabilizer eigenstates, offering insight into their potential for controllable state preparation in constrained
quantum systems. For a qubit layout as shown in Fig.~\ref{fig:qubit_layout}, the stabilizer scar state of Fig. \ref{fig:2x2pbc} $(i)$ with PBC, can be written in Fock basis as, 
\[
\ket{\psi^-} = \frac{1}{\sqrt{2}} \left( \ket{b_0} - \ket{b_1} \right) =  \frac{1}{\sqrt{2}}\left( \ket{00110110} - \ket{11001001} \right),
\]
where we follow the convention $\ket{q_7,\dots,q_0}$.
This 8 qubit state is prepared using a gate sequence shown in Fig. \ref{fig:circuit}. The circuit begins by initializing the system in the computational basis state $\ket{b_0}$ using a layer of $X$ gates on the appropriate qubits. We then identify the subset of qubits $\mathcal{S}$ where the configurations $\ket{b_0}$ and $\ket{b_1}$ differ. These are precisely the links affected by the action of a local plaquette-flip operator. To coherently generate the superposition, we apply a Hadamard gate to a single pivot qubit $q_p \in \mathcal{S}$ followed by a sequence of CNOT gates from $q_p$ to all other qubits in $\mathcal{D}$. This operation entangles the computational branches to create the symmetric state $(\ket{b_0} + \ket{b_1})/\sqrt{2}$. A single $Z$ gate on the pivot introduces a relative minus sign between the branches, yielding the antisymmetric eigenstate $\ket{\psi^-}$.

 

This construction applies to any pair of configurations related by a local plaquette flip and satisfying Gauss's law. The procedure uses only Clifford gates, and the number of gates depends solely on the size of the plaquette support, not on the overall system size. As a result, the circuit has constant depth per plaquette-pair singlet. When extended across a larger lattice, multiple copies of the local circuit can be applied in parallel on disjoint plaquette pairs to construct the stablizer sublattice scar state. 

The circuit depth scales linearly with the number of plaquette pairs, and all operations are composed solely of single-qubit rotations and two-qubit CNOT gates, rendering the protocol experimentally feasible on current NISQ hardware.



\section{Conclusions and Outlook}
\label{sec:conclusions}

The RK model, a paradigmatic and physically local Hamiltonian of lattice gauge theory, hosts exact stabilizer-scarred eigenstates embedded within its otherwise nonintegrable spectrum.
By combining exact diagonalization with stabilizer-based and multifractal diagnostics, we have identified a distinct family of nonthermal eigenstates, sublattice stabilizer scars, that simultaneously violate ETH and possess a stabilizer structure.
These scars are characterized by vanishing SRE, finite subthermal bipartite entanglement, and integer eigenvalues of both the kinetic and potential operators, reflecting an emergent ordered pattern across the checkerboard sublattice. This sublattice order persists across all system sizes studied.

Despite arising in a non-commuting, non-integrable Hamiltonian rather than a commuting stabilizer construction, these states admit an exact canonical stabilizer representation and can be efficiently prepared using Clifford circuits. While the RK model is known to host a rich variety of scarred eigenstates~\cite{biswas2022scars}, our results indicate that sublattice scars form a distinguished subclass that admits a stabilizer description. Although our analysis does not exclude the possibility of stabilizer scars of other types in dimensions $L_x,L_y\ge 4$, it strongly suggests that the emergence of a stabilizer structure is intimately tied to the sublattice nature of the scars.

Our results establish a direct connection between stabilizer quantum information theory and quantum many-body scarring in a physically realizable model. 
We demonstrate that stabilizer-protected subspaces need not be engineered through artificial commuting-projector Hamiltonians, but can instead emerge intrinsically in realistic gauge-theoretic systems. 
This finding opens a new perspective on the interplay between locality, constrained dynamics, and classical simulability in strongly correlated quantum matter, and suggests promising directions for experimentally probing stabilizer scars using programmable quantum simulators and near-term quantum processors. This connection suggests that certain sectors of gauge theories can host
stabilizer-protected subspaces that remain isolated from thermalization,
providing a natural platform for robust quantum state engineering. 

Future work can explore the stability of stabilizer scars under generic gauge-invariant perturbations and determine whether their stabilizer structure persists approximately beyond the Rokhsar--Kivelson point, potentially giving rise to long-lived or prethermal nonthermal manifolds.  Investigating their dynamical signatures, including coherent revivals and constrained evolution within the stabilizer codespace, will be essential for identifying experimentally observable consequences. Quantum machine learning methods~\cite{Szoldra22, Cao2024, Feng2025} can be useful tools in such explorations. Extending this framework to non-Abelian lattice gauge theories may reveal richer stabilizer structures and connections to topological order and logical encoding in gauge-invariant systems. More broadly, the stabilizer perspective suggests a natural interpretation of scarred subspaces as emergent quantum error-correcting codes embedded in physical many-body systems. Clarifying the relationship between gauge constraints, Hilbert-space fragmentation, and stabilizer structure may provide a unified framework linking quantum information structure, classical simulability, and the emergence of nonthermal behavior in constrained quantum matter.

\section*{Data Availability Statement}
\label{sec:data_availability}
The data that support the findings of this study are available from the corresponding author upon request.
\section*{Acknowledgments}
\label{sec:acknowledgments}
P.Stornati would like to thank Debasish Banerjee, Egle Pagliaro, and Emanuele Tirrito for helpful discussions on related topics. 
S.G. and L.S. acknowledge the support of the Deutsche Forschungsgemeinschaft (DFG, German Research Foundation) under Germany's Excellence Strategy -- EXC-2123 Quantum-Frontiers -- 390837967.
P.S. and P.S. acknowledges funding from the Spanish Ministry for Digital Transformation and the Civil Service of the Spanish Government through the QUANTUM ENIA project call - Quantum Spain, EU, through the Recovery, Transformation and Resilience Plan – NextGenerationEU, within the framework of Digital Spain 2026.

\bibliographystyle{apsrev4-2}  
\bibliography{references}

@article{PhysRevA.110.062427,
  title = {Doped stabilizer states in many-body physics and where to find them},
  author = {Gu, Andi and Oliviero, Salvatore F. E. and Leone, Lorenzo},
  journal = {Phys. Rev. A},
  volume = {110},
  issue = {6},
  pages = {062427},
  numpages = {7},
  year = {2024},
  month = {Dec},
  publisher = {American Physical Society},
  doi = {10.1103/PhysRevA.110.062427},
  url = {https://link.aps.org/doi/10.1103/PhysRevA.110.062427}
}

@article{Banerjee_2013,
   title={The (2 + 1)-d U(1) quantum link model masquerading as deconfined criticality},
   volume={2013},
   ISSN={1742-5468},
   url={http://dx.doi.org/10.1088/1742-5468/2013/12/P12010},
   DOI={10.1088/1742-5468/2013/12/p12010},
   number={12},
   journal={Journal of Statistical Mechanics: Theory and Experiment},
   publisher={IOP Publishing},
   author={Banerjee, D and Jiang, F-J and Widmer, P and Wiese, U-J},
   year={2013},
   month=dec, pages={P12010} }

@article{PhysRevLett.126.220601,
  title = {Quantum Scars from Zero Modes in an Abelian Lattice Gauge Theory on Ladders},
  author = {Banerjee, Debasish and Sen, Arnab},
  journal = {Phys. Rev. Lett.},
  volume = {126},
  issue = {22},
  pages = {220601},
  numpages = {6},
  year = {2021},
  month = {Jun},
  publisher = {American Physical Society},
  doi = {10.1103/PhysRevLett.126.220601},
  url = {https://link.aps.org/doi/10.1103/PhysRevLett.126.220601}
}

@article{Stornati_2023,
   title={Crystalline phases at finite winding densities in a quantum link ladder},
   volume={107},
   ISSN={2470-0029},
   url={http://dx.doi.org/10.1103/PhysRevD.107.L031504},
   DOI={10.1103/physrevd.107.l031504},
   number={3},
   journal={Physical Review D},
   publisher={American Physical Society (APS)},
   author={Stornati, Paolo and Krah, Philipp and Jansen, Karl and Banerjee, Debasish},
   year={2023},
   month=feb }

@article{sublatticescars2024,
  title = {Sublattice scars and beyond in two-dimensional $U(1)$ quantum link lattice gauge theories},
  author = {Sau, Indrajit and Stornati, Paolo and Banerjee, Debasish and Sen, Arnab},
  journal = {Phys. Rev. D},
  volume = {109},
  issue = {3},
  pages = {034519},
  numpages = {17},
  year = {2024},
  month = {Feb},
  publisher = {American Physical Society},
  doi = {10.1103/PhysRevD.109.034519},
  url = {https://link.aps.org/doi/10.1103/PhysRevD.109.034519}
}

@article{PhysRevD.110.094506,
  title = {Quantum many-body scars for arbitrary integer spin in $2+1\mathrm{D}$ Abelian gauge theories},
  author = {Budde, Thea and Krstic Marinkovic, Marina and Pinto Barros, Joao C.},
  journal = {Phys. Rev. D},
  volume = {110},
  issue = {9},
  pages = {094506},
  numpages = {9},
  year = {2024},
  month = {Nov},
  publisher = {American Physical Society},
  doi = {10.1103/PhysRevD.110.094506},
  url = {https://link.aps.org/doi/10.1103/PhysRevD.110.094506},
}

@article{ScarsSchwingerPhysRevB.107.205112,
  title = {Prominent quantum many-body scars in a truncated Schwinger model},
  author = {Desaules, Jean-Yves and Hudomal, Ana and Banerjee, Debasish and Sen, Arnab and Papi\ifmmode \acute{c}\else \'{c}\fi{}, Zlatko and Halimeh, Jad C.},
  journal = {Phys. Rev. B},
  volume = {107},
  issue = {20},
  pages = {205112},
  numpages = {21},
  year = {2023},
  month = {May},
  publisher = {American Physical Society},
  doi = {10.1103/PhysRevB.107.205112},
  url = {https://link.aps.org/doi/10.1103/PhysRevB.107.205112}
}

@article{lan2017eth,
  title={Eigenstate thermalization hypothesis in quantum dimer models},
  author={Lan, Tian and Powell, Stephen},
  journal={Physical Review B},
  volume={96},
  number={11},
  pages={115140},
  year={2017},
  publisher={APS},
  doi={10.1103/PhysRevB.96.115140},
  url = {https://link.aps.org/doi/10.1103/PhysRevB.96.115140},

}

@article{banerjee2021scars,
  title={Quantum scars from zero modes in an abelian lattice gauge theory on ladders},
  author={Banerjee, Debasish and Sen, Arnab},
  journal={Physical Review Letters},
  volume={126},
  number={22},
  pages={220601},
  year={2021},
  publisher={APS},
  doi={10.1103/PhysRevLett.126.220601},
  url = {https://link.aps.org/doi/10.1103/PhysRevLett.126.220601},
}

@article{ScarsPXPPhysRevLett.134.160401,
  title = {Quantum Many-Body Scars beyond the PXP Model in Rydberg Simulators},
  author = {Kerschbaumer, Aron and Ljubotina, Marko and Serbyn, Maksym and Desaules, Jean-Yves},
  journal = {Phys. Rev. Lett.},
  volume = {134},
  issue = {16},
  pages = {160401},
  numpages = {9},
  year = {2025},
  month = {Apr},
  publisher = {American Physical Society},
  doi = {10.1103/PhysRevLett.134.160401},
  url = {https://link.aps.org/doi/10.1103/PhysRevLett.134.160401}
}

@article{Halimeh2023robustquantummany,
  doi = {10.22331/q-2023-05-15-1004},
  url = {https://doi.org/10.22331/q-2023-05-15-1004},
  title = {Robust quantum many-body scars in lattice gauge theories},
  author = {Halimeh, Jad C. and Barbiero, Luca and Hauke, Philipp and Grusdt, Fabian and Bohrdt, Annabelle},
  journal = {{Quantum}},
  issn = {2521-327X},
  publisher = {{Verein zur F{\"{o}}rderung des Open Access Publizierens in den Quantenwissenschaften}},
  volume = {7},
  pages = {1004},
  month = may,
  year = {2023}
}

@article{ExactQMBSPhysRevB.108.195133,
  title = {Exact quantum many-body scars in higher-spin kinetically constrained models},
  author = {Yuan, Dong and Zhang, Shun-Yao and Deng, Dong-Ling},
  journal = {Phys. Rev. B},
  volume = {108},
  issue = {19},
  pages = {195133},
  numpages = {10},
  year = {2023},
  month = {Nov},
  publisher = {American Physical Society},
  doi = {10.1103/PhysRevB.108.195133},
  url = {https://link.aps.org/doi/10.1103/PhysRevB.108.195133}
}

@article{QMBS1dPhysRevLett.132.230403,
  title = {Nonmesonic Quantum Many-Body Scars in a 1D Lattice Gauge Theory},
  author = {Ge, Zi-Yong and Zhang, Yu-Ran and Nori, Franco},
  journal = {Phys. Rev. Lett.},
  volume = {132},
  issue = {23},
  pages = {230403},
  numpages = {8},
  year = {2024},
  month = {Jun},
  publisher = {American Physical Society},
  doi = {10.1103/PhysRevLett.132.230403},
  url = {https://link.aps.org/doi/10.1103/PhysRevLett.132.230403}
}

@article{biswas2022scars,
  title={Scars from protected zero modes and beyond in U(1) quantum link and quantum dimer models},
  author={Biswas, Soumyadeep and Banerjee, Debasish and Sen, Arnab},
  journal={SciPost Physics},
  volume={12},
  pages={148},
  year={2022},
  doi={10.21468/SciPostPhys.12.5.148},
url={https://scipost.org/10.21468/SciPostPhys.12.5.148},
}

@article{leone2022stabilizer,
  title={Stabilizer Rényi entropy},
  author={Leone, Lorenzo and Oliviero, Salvatore F E and Hamma, Alioscia},
  journal={Physical Review Letters},
  volume={128},
  number={5},
  pages={050402},
  year={2022},
  publisher={APS},
  doi={10.1103/PhysRevLett.128.050402},
url = {https://link.aps.org/doi/10.1103/PhysRevLett.128.050402}
}

@article{rokhsar1988superconductivity,
  title={Superconductivity and the quantum hard-core dimer gas},
  author={Rokhsar, Daniel S and Kivelson, Steven A},
  journal={Physical Review Letters},
  volume={61},
  number={20},
  pages={2376--2379},
  year={1988},
  publisher={APS},
  doi={10.1103/PhysRevLett.61.2376},
url = {https://link.aps.org/doi/10.1103/PhysRevLett.61.2376}


}

@article{MultifractalFlatness2023,
  title = {Measuring nonstabilizerness via multifractal flatness},
  author = {Turkeshi, Xhek and Schir\`o, Marco and Sierant, Piotr},
  journal = {Phys. Rev. A},
  volume = {108},
  issue = {4},
  pages = {042408},
  numpages = {8},
  year = {2023},
  month = {Oct},
  publisher = {American Physical Society},
  doi = {10.1103/PhysRevA.108.042408},
  url = {https://link.aps.org/doi/10.1103/PhysRevA.108.042408}
}

@article{Deutsch1991,
  title = {Quantum statistical mechanics in a closed system},
  author = {Deutsch, J. M.},
  journal = {Phys. Rev. A},
  volume = {43},
  issue = {4},
  pages = {2046--2049},
  numpages = {0},
  year = {1991},
  month = {Feb},
  publisher = {American Physical Society},
  doi = {10.1103/PhysRevA.43.2046},
  url = {https://link.aps.org/doi/10.1103/PhysRevA.43.2046}
}

@article{Srednicki1994,
  title = {Chaos and quantum thermalization},
  author = {Srednicki, Mark},
  journal = {Phys. Rev. E},
  volume = {50},
  issue = {2},
  pages = {888--901},
  numpages = {0},
  year = {1994},
  month = {Aug},
  publisher = {American Physical Society},
  doi = {10.1103/PhysRevE.50.888},
  url = {https://link.aps.org/doi/10.1103/PhysRevE.50.888}
}

@article{DAlessio2016,
author = {Luca D'Alessio and Yariv Kafri and Anatoli Polkovnikov and Marcos Rigol},
title = {From quantum chaos and eigenstate thermalization to statistical mechanics and thermodynamics},
journal = {Advances in Physics},
volume = {65},
number = {3},
pages = {239--362},
year = {2016},
publisher = {Taylor \& Francis},
doi = {10.1080/00018732.2016.1198134},
URL = {https://doi.org/10.1080/00018732.2016.1198134},


}

@article{Turner2018,
author={Turner, C. J.
and Michailidis, A. A.
and Abanin, D. A.
and Serbyn, M.
and Papi{\'{c}}, Z.},
title={Weak ergodicity breaking from quantum many-body scars},
journal={Nature Physics},
year={2018},
month={Jul},
day={01},
volume={14},
number={7},
pages={745-749},
issn={1745-2481},
doi={10.1038/s41567-018-0137-5},
url={https://doi.org/10.1038/s41567-018-0137-5}
}

@article{Serbyn2021,
author={Serbyn, Maksym
and Abanin, Dmitry A.
and Papi{\'{c}}, Zlatko},
title={Quantum many-body scars and weak breaking of ergodicity},
journal={Nature Physics},
year={2021},
month={Jun},
day={01},
volume={17},
number={6},
pages={675-685},
issn={1745-2481},
doi={10.1038/s41567-021-01230-2},
url={https://doi.org/10.1038/s41567-021-01230-2}
}

@article{Bernien2017,
author={Bernien, Hannes
and Schwartz, Sylvain
and Keesling, Alexander
and Levine, Harry
and Omran, Ahmed
and Pichler, Hannes
and Choi, Soonwon
and Zibrov, Alexander S.
and Endres, Manuel
and Greiner, Markus
and Vuleti{\'{c}}, Vladan
and Lukin, Mikhail D.},
title={Probing many-body dynamics on a 51-atom quantum simulator},
journal={Nature},
year={2017},
month={Nov},
day={01},
volume={551},
number={7682},
pages={579-584},
issn={1476-4687},
doi={10.1038/nature24622},
url={https://doi.org/10.1038/nature24622}
}

@article{Ho2019,
  title = {Periodic Orbits, Entanglement, and Quantum Many-Body Scars in Constrained Models: Matrix Product State Approach},
  author = {Ho, Wen Wei and Choi, Soonwon and Pichler, Hannes and Lukin, Mikhail D.},
  journal = {Phys. Rev. Lett.},
  volume = {122},
  issue = {4},
  pages = {040603},
  numpages = {6},
  year = {2019},
  month = {Jan},
  publisher = {American Physical Society},
  doi = {10.1103/PhysRevLett.122.040603},
  url = {https://link.aps.org/doi/10.1103/PhysRevLett.122.040603}
}

@article{Surace2020,
  title = {Lattice Gauge Theories and String Dynamics in Rydberg Atom Quantum Simulators},
  author = {Surace, Federica M. and Mazza, Paolo P. and Giudici, Giuliano and Lerose, Alessio and Gambassi, Andrea and Dalmonte, Marcello},
  journal = {Phys. Rev. X},
  volume = {10},
  issue = {2},
  pages = {021041},
  numpages = {14},
  year = {2020},
  month = {May},
  publisher = {American Physical Society},
  doi = {10.1103/PhysRevX.10.021041},
  url = {https://link.aps.org/doi/10.1103/PhysRevX.10.021041}
}

@article{Hamma05,
  title = {Bipartite entanglement and entropic boundary law in lattice spin systems},
  author = {Hamma, Alioscia and Ionicioiu, Radu and Zanardi, Paolo},
  journal = {Phys. Rev. A},
  volume = {71},
  issue = {2},
  pages = {022315},
  numpages = {10},
  year = {2005},
  month = {Feb},
  publisher = {American Physical Society},
  doi = {10.1103/PhysRevA.71.022315},
  url = {https://link.aps.org/doi/10.1103/PhysRevA.71.022315}
}

@article{Sala2020,
  title = {Ergodicity Breaking Arising from Hilbert Space Fragmentation in Dipole-Conserving Hamiltonians},
  author = {Sala, Pablo and Rakovszky, Tibor and Verresen, Ruben and Knap, Michael and Pollmann, Frank},
  journal = {Phys. Rev. X},
  volume = {10},
  issue = {1},
  pages = {011047},
  numpages = {19},
  year = {2020},
  month = {Feb},
  publisher = {American Physical Society},
  doi = {10.1103/PhysRevX.10.011047},
  url = {https://link.aps.org/doi/10.1103/PhysRevX.10.011047}
}

@article{Dehaene03,
  title = {Clifford group, stabilizer states, and linear and quadratic operations over GF(2)},
  author = {Dehaene, Jeroen and De Moor, Bart},
  journal = {Phys. Rev. A},
  volume = {68},
  issue = {4},
  pages = {042318},
  numpages = {10},
  year = {2003},
  month = {Oct},
  publisher = {American Physical Society},
  doi = {10.1103/PhysRevA.68.042318},
  url = {https://link.aps.org/doi/10.1103/PhysRevA.68.042318}
}

@article{Tirrito24,
  title = {Quantifying nonstabilizerness through entanglement spectrum flatness},
  author = {Tirrito, Emanuele and Tarabunga, Poetri Sonya and Lami, Gugliemo and Chanda, Titas and Leone, Lorenzo and Oliviero, Salvatore F. E. and Dalmonte, Marcello and Collura, Mario and Hamma, Alioscia},
  journal = {Phys. Rev. A},
  volume = {109},
  issue = {4},
  pages = {L040401},
  numpages = {6},
  year = {2024},
  month = {Apr},
  publisher = {American Physical Society},
  doi = {10.1103/PhysRevA.109.L040401},
  url = {https://link.aps.org/doi/10.1103/PhysRevA.109.L040401}
}

@article{Sierant22multi,
  title = {Universal Behavior beyond Multifractality of Wave Functions at Measurement-Induced Phase Transitions},
  author = {Sierant, Piotr and Turkeshi, Xhek},
  journal = {Phys. Rev. Lett.},
  volume = {128},
  issue = {13},
  pages = {130605},
  numpages = {7},
  year = {2022},
  month = {Apr},
  publisher = {American Physical Society},
  doi = {10.1103/PhysRevLett.128.130605},
  url = {https://link.aps.org/doi/10.1103/PhysRevLett.128.130605}
}

@misc{Sierant26computingSRE,
      title={Computing quantum magic of state vectors}, 
      author={Piotr Sierant and Jofre Vallès-Muns and Artur Garcia-Saez},
      year={2026},
      eprint={2601.07824},
      archivePrefix={arXiv},
      primaryClass={quant-ph},
      url={https://arxiv.org/abs/2601.07824}, 
}

@article{Szoldra22,
  title = {Unsupervised detection of decoupled subspaces: Many-body scars and beyond},
  author = {Szo\l{}dra, Tomasz and Sierant, Piotr and Lewenstein, Maciej and Zakrzewski, Jakub},
  journal = {Phys. Rev. B},
  volume = {105},
  issue = {22},
  pages = {224205},
  numpages = {7},
  year = {2022},
  month = {Jun},
  publisher = {American Physical Society},
  doi = {10.1103/PhysRevB.105.224205},
  url = {https://link.aps.org/doi/10.1103/PhysRevB.105.224205}
}

@article{Cao2024,
doi = {10.1088/2632-2153/ad4d3f},
url = {https://doi.org/10.1088/2632-2153/ad4d3f},
year = {2024},
month = {may},
publisher = {IOP Publishing},
volume = {5},
number = {2},
pages = {025049},
author = {Cao, Harvey and Angelakis, Dimitris G and Leykam, Daniel},
title = {Unsupervised learning of quantum many-body scars using intrinsic dimension},
journal = {Machine Learning: Science and Technology}
}

@Article{Feng2025,
author={Feng, Jia-Jin
and Zhang, Bingzhi
and Yang, Zhi-Cheng
and Zhuang, Quntao},
title={Uncovering quantum many-body scars with quantum machine learning},
journal={npj Quantum Information},
year={2025},
month={Mar},
day={11},
volume={11},
number={1},
pages={42},
issn={2056-6387},
doi={10.1038/s41534-025-01005-0},
url={https://doi.org/10.1038/s41534-025-01005-0}
}

@article{Amico08entanglement,
  title = {Entanglement in many-body systems},
  author = {Amico, Luigi and Fazio, Rosario and Osterloh, Andreas and Vedral, Vlatko},
  journal = {Rev. Mod. Phys.},
  volume = {80},
  issue = {2},
  pages = {517--576},
  numpages = {0},
  year = {2008},
  month = {May},
  publisher = {American Physical Society},
  doi = {10.1103/RevModPhys.80.517},
  url = {https://link.aps.org/doi/10.1103/RevModPhys.80.517}
}

@article{Horodecki09quantum,
  title = {Quantum entanglement},
  author = {Horodecki, Ryszard and Horodecki, Pawe\l{} and Horodecki, Micha\l{} and Horodecki, Karol},
  journal = {Rev. Mod. Phys.},
  volume = {81},
  issue = {2},
  pages = {865--942},
  numpages = {0},
  year = {2009},
  month = {Jun},
  publisher = {American Physical Society},
  doi = {10.1103/RevModPhys.81.865},
  url = {https://link.aps.org/doi/10.1103/RevModPhys.81.865}
}

@article{Shannon12,
  title = {Quantum Ice: A Quantum Monte Carlo Study},
  author = {Shannon, Nic and Sikora, Olga and Pollmann, Frank and Penc, Karlo and Fulde, Peter},
  journal = {Phys. Rev. Lett.},
  volume = {108},
  issue = {6},
  pages = {067204},
  numpages = {5},
  year = {2012},
  month = {Feb},
  publisher = {American Physical Society},
  doi = {10.1103/PhysRevLett.108.067204},
  url = {https://link.aps.org/doi/10.1103/PhysRevLett.108.067204}
}

@article{Hermele04,
  title = {Pyrochlore photons: The $U(1)$ spin liquid in a $S=\frac{1}{2}$ three-dimensional frustrated magnet},
  author = {Hermele, Michael and Fisher, Matthew P. A. and Balents, Leon},
  journal = {Phys. Rev. B},
  volume = {69},
  issue = {6},
  pages = {064404},
  numpages = {21},
  year = {2004},
  month = {Feb},
  publisher = {American Physical Society},
  doi = {10.1103/PhysRevB.69.064404},
  url = {https://link.aps.org/doi/10.1103/PhysRevB.69.064404}
}

@article{Khemani2020,
  title = {Localization from Hilbert space shattering: From theory to physical realizations},
  author = {Khemani, Vedika and Hermele, Michael and Nandkishore, Rahul},
  journal = {Phys. Rev. B},
  volume = {101},
  issue = {17},
  pages = {174204},
  numpages = {17},
  year = {2020},
  month = {May},
  publisher = {American Physical Society},
  doi = {10.1103/PhysRevB.101.174204},
  url = {https://link.aps.org/doi/10.1103/PhysRevB.101.174204}
}

@article{Brenes18,
  title = {Many-Body Localization Dynamics from Gauge Invariance},
  author = {Brenes, Marlon and Dalmonte, Marcello and Heyl, Markus and Scardicchio, Antonello},
  journal = {Phys. Rev. Lett.},
  volume = {120},
  issue = {3},
  pages = {030601},
  numpages = {6},
  year = {2018},
  month = {Jan},
  publisher = {American Physical Society},
  doi = {10.1103/PhysRevLett.120.030601},
  url = {https://link.aps.org/doi/10.1103/PhysRevLett.120.030601}
}

@article{Moudgalya2022,
doi = {10.1088/1361-6633/ac73a0},
url = {https://doi.org/10.1088/1361-6633/ac73a0},
year = {2022},
month = {jul},
publisher = {IOP Publishing},
volume = {85},
number = {8},
pages = {086501},
author = {Moudgalya, Sanjay and Bernevig, B Andrei and Regnault, Nicolas},
title = {Quantum many-body scars and Hilbert space fragmentation: a review of exact results},
journal = {Reports on Progress in Physics}
}

@article{Moessner2001,
  title = {Resonating Valence Bond Phase in the Triangular Lattice Quantum Dimer Model},
  author = {Moessner, R. and Sondhi, S. L.},
  journal = {Phys. Rev. Lett.},
  volume = {86},
  issue = {9},
  pages = {1881--1884},
  numpages = {0},
  year = {2001},
  month = {Feb},
  publisher = {American Physical Society},
  doi = {10.1103/PhysRevLett.86.1881},
  url = {https://link.aps.org/doi/10.1103/PhysRevLett.86.1881}
}

@article{Banerjee2014,

  title = {Atomic Quantum Simulation of Dynamical Gauge Fields Coupled to Fermionic Matter: From String Breaking to Evolution after a Quench},
  author = {Banerjee, D. and Dalmonte, M. and M\"uller, M. and Rico, E. and Stebler, P. and Wiese, U.-J. and Zoller, P.},
  journal = {Phys. Rev. Lett.},
  volume = {109},
  issue = {17},
  pages = {175302},
  numpages = {5},
  year = {2012},
  month = {Oct},
  publisher = {American Physical Society},
  doi = {10.1103/PhysRevLett.109.175302},
  url = {https://link.aps.org/doi/10.1103/PhysRevLett.109.175302}
}

@article{Chandran2023,
   author = "Chandran, Anushya and Iadecola, Thomas and Khemani, Vedika and Moessner, Roderich",
   title = "Quantum Many-Body Scars: A Quasiparticle Perspective", 
   journal= "Annual Review of Condensed Matter Physics",
   year = "2023",
   volume = "14",
   number = "Volume 14, 2023",
   pages = "443-469",
   doi = "https://doi.org/10.1146/annurev-conmatphys-031620-101617",
   url = "https://www.annualreviews.org/content/journals/10.1146/annurev-conmatphys-031620-101617",
   publisher = "Annual Reviews",
   issn = "1947-5462",
   type = "Journal Article",
  }

@article{Stabilizerscars,
  title = {Stabilizer Scars},
  author = {Hartse, Jeremy and Fidkowski, Lukasz and Mueller, Niklas},
  journal = {Phys. Rev. Lett.},
  volume = {135},
  issue = {6},
  pages = {060402},
  numpages = {9},
  year = {2025},
  month = {Aug},
  publisher = {American Physical Society},
  doi = {10.1103/n5hb-l5p5},
  url = {https://link.aps.org/doi/10.1103/n5hb-l5p5}
}

@article{StabilizerCircuitsPhysRevA.70.052328,
  title = {Improved simulation of stabilizer circuits},
  author = {Aaronson, Scott and Gottesman, Daniel},
  journal = {Phys. Rev. A},
  volume = {70},
  issue = {5},
  pages = {052328},
  numpages = {14},
  year = {2004},
  month = {Nov},
  publisher = {American Physical Society},
  doi = {10.1103/PhysRevA.70.052328},
  url = {https://link.aps.org/doi/10.1103/PhysRevA.70.052328}
}

@misc{montanaro2017learningstabilizerstatesbell,
      title={Learning stabilizer states by Bell sampling}, 
      author={Ashley Montanaro},
      year={2017},
      eprint={1707.04012},
      archivePrefix={arXiv},
      primaryClass={quant-ph},
      url={https://arxiv.org/abs/1707.04012}, 
}

@article{Leone24,
  title = {Stabilizer entropies are monotones for magic-state resource theory},
  author = {Leone, Lorenzo and Bittel, Lennart},
  journal = {Phys. Rev. A},
  volume = {110},
  issue = {4},
  pages = {L040403},
  numpages = {6},
  year = {2024},
  month = {Oct},
  publisher = {American Physical Society},
  doi = {10.1103/PhysRevA.110.L040403},
  url = {https://link.aps.org/doi/10.1103/PhysRevA.110.L040403}
}

@misc{dooley2026parenthamiltoniansstabilizerquantum,
      title={Parent Hamiltonians for stabilizer quantum many-body scars}, 
      author={Shane Dooley},
      year={2026},
      eprint={2601.10805},
      archivePrefix={arXiv},
      primaryClass={quant-ph},
      url={https://arxiv.org/abs/2601.10805}, 
}

@article{KITAEV20032,
title = {Fault-tolerant quantum computation by anyons},
journal = {Annals of Physics},
volume = {303},
number = {1},
pages = {2-30},
year = {2003},
issn = {0003-4916},
doi = {https://doi.org/10.1016/S0003-4916(02)00018-0},
url = {https://www.sciencedirect.com/science/article/pii/S0003491602000180},
author = {A.Yu. Kitaev}
}
\end{document}